\documentstyle[12pt]{article}

\def\8{\infty}
\def\oh{\frac{1}{2}}

\def\tt{\frac{2}{3}}

\def\d{\partial}
\def\i{\imath\,}

\def\dal{\partial_{\alpha}}
\def\dbe{\partial_{\beta}}

\def\undertext#1{\vtop{\hbox{#1}\kern 1pt \hrule}}
\def\ra{\rightarrow}
\def\Ra{\Rightarrow}

\def\VEV#1{\langle #1\rangle}

\def\dd#1{\frac{d}{d#1}}

\def\pp#1{\frac{\partial}{\partial#1}}

\def\ff#1{\frac{\delta}{\delta#1}}
\def\fbyf#1#2{\frac{\delta#1}{\delta#2}}

\def\br{\\ \nonumber & &}
\def\inv#1{\frac{1}{#1}}
\def\be{\begin{equation}}
\def\ee{\end{equation}}
\def\bea{\begin{eqnarray} & &}
\def\eea{\end{eqnarray}}
\def\ct#1{\cite{#1}}
\def\rf#1{(\ref{#1})}
\def\EXP#1{\exp\left(#1\right)}

\def\PBR#1{\left[#1\right]}
\def\JAC#1{\frac{\d\left(#1\right)}{\d\left(\rho_1,
\rho_2,\rho_3\right)}}

\def\C{\oint_C d }
\def\Y{\Psi_C(\gamma)}

\def\PB{Poisson brackets }

\def\dt{\partial_t\,}
\def\val{v_{\alpha}}
\def\vbe{v_{\beta}}
\def\vga{v_{\gamma}}

\def\ral{r_{\alpha}}
\def\rbe{r_{\beta}}

\begin{document}

\begin{titlepage}

{\bf June '93}\hfill	  {\bf PUPT-1409}\\

\begin{center}

{\bf  TURBULENCE AS STATISTICS OF VORTEX CELLS}

\vspace{1.5cm}

{\bf   A.A.~Migdal}

\vspace{1.0cm}

{\it  Physics Department, Princeton University,\\
Jadwin Hall, Princeton, NJ 08544-1000.\\
E-mail: migdal@acm.princeton.edu}

\vspace{1.9cm}

\end{center}

\abstract{We develop the formulation of turbulence in terms of  the
functional
integral over the  phase space configurations of  the vortex cells.
The phase
space consists of Clebsch coordinates at the surface of the  vortex
cells plus
the Lagrange coordinates of this surface plus the conformal metric.
Using the
Hamiltonian dynamics we find an invariant probability distribution
which
satisfies the Liouville equation. The violations of  the time
reversal
invariance come from certain topological terms in effective energy of
our
Gibbs-like distribution. We study the topological aspects of the
statistics and
use the string theory methods to estimate intermittency.
}
\vfill
\end{titlepage}

\tableofcontents
\newpage
\section{Introduction}

The central problem of turbulence is to find the analog of the Gibbs
distribution for the energy cascade. The mathematical formulation of
this
problem is amazingly simple. In inertial range we could neglect
the viscosity and forcing and study the Euler  dynamics of ideal
incompressible fluid
\begin{equation}\label{EL}
\dt \val = \vbe \omega_{\alpha \beta} - \dal  h
\end{equation}
Here
\begin{equation}\label{OM}
\omega_{\alpha \beta} = \dal \vbe - \dbe \val
\end{equation}
is the vorticity field and
\begin{equation}\label{H}
h = p + \oh \val^2
\end{equation}
is the enthalpy which is eliminated from the incompressibility
condition
\begin{equation}\label{dv}
\dal \val = 0
\end{equation}

The key point is that the Euler  dynamics can be regarded as a
Hamiltonian flow
in functional phase space. This geometric view was first proposed by
Arnold in
1966 (see his famous book \ct{Arn}) and later developed by other
mathematicians
(see the references in the Moffatt's lecture in the 1992 Santa
Barbara
conference proceedings\ct{SBconf}). Here we derive the Hamiltonian
dynamics
from scratch using the conventional physical terminology. All the
necessary
computations are presented in Appendix A. Some of these results are
new.

The Hamiltonian here is just the fluid energy
\be
H = \int d^3 r \oh \val^2
\ee
and the phase space corresponds to all the velocity fields subject to
the
incompressibility constraint. The \PB between the components of
velocity field
\be
\PBR{\val(r_1) ,\vbe(r_2)} = \int d^3 r
T_{\alpha\mu}\left(r_1-r\right)
 T_{\beta\nu}\left(r_2-r\right)\omega_{\mu\nu}(r)
\ee
where
\be
T_{\alpha\beta}(r) = \delta_{\alpha\beta} \delta(r) + \dal\dbe
\frac{1}{4\pi r}
\ee
is the projection operator. The Euler equation can be written in a
manifestly
Hamiltonian form
\be
\d_t\val = \int d^3 r'
T_{\alpha\beta}\left(r-r'\right)  \omega_{\beta\gamma}(r') \vga(r')=
\PBR{\val, H}
\ee

The Liouville theorem of the phase space volume  conservation applies
here
\begin{equation}\label{Li1}
\ff{\val(r)} \PBR{\val(r), H}=0
\end{equation}
\be
(Dv) = \prod_r d^3v(r) =\mbox{ const}
\ee
The probability distributions $ P[v] $ compatible with this dynamics,
must also
be conserved, i.e. it must commute with the Hamiltonian
\begin{equation}\label{Li2}
\PBR{ P[v], H}  = 0
\end{equation}

The Gibbs distribution
\begin{equation}\label{Gb}
 P[v] = \EXP{- \beta H }
\end{equation}
is the only known general solution of the Liouville equation
(\ref{Li2}). The
above mentioned central problem of turbulence is to find another one.

This formulation of turbulence is significantly different from the
popular
formulation based on the so called Wyld diagram technique \ct{Lv91}.
There, the
functional integral is in place from the very beginning but the
problem is to
find its turbulent limit (zero viscosity). This functional integral
involves
time, so it describes the kinetic phenomena in addition to the steady
state we
are studying here.

After trying for few years to do something with the Wyld approach  I
conclude
that this is a dead end. The best bet here would be the
renormalization group,
which magically works in statistical physics. Those critical
phenomena were
close to Gaussian, which allowed Wilson to develop the $ \epsilon $
expansion
by rearranging  the ordinary perturbation expansion.

There is no such luck in turbulence. The nonlinear effects are much
stronger.
The  observed variety of vorticity structures with their long range
interactions does not look  like  the block spins of critical
phenomena.
Moreover, there are notorious infrared divergencies, which make
problematic the
whole existence of the universal kinetics of turbulence. No! These
old tricks
are not going to work, we have to invent the new ones.

If we are looking for something pure and simple this might be the
steady state
distribution of vorticity structures. Here the geometric  methods may
allow us
to go much further than  the general methods of quantum field theory.
Being
regarded as a problem of fractal geometry rather than a nonlinear
wave problem,
turbulence may reveal some mathematical beauty to match the beauty of
the
Euler-Lagrange dynamics.

This dream  motivated the geometric approach to the vortex sheet
dynamics in
\ct{AMi}, where the attempt was made to simulate turbulence as the
stochastic
motion of the vortex sheet. This project ran out of computer
resources, as  it
happened before to many other projects of that kind. However, the
geometric
tools developed in that work proved to be useful and we are going to
use them
here.

Another false start: I tried to solve the Liouville equation
variationally,
with the Gaussian Anzatz with anomalous dimensions for the velocity
field.\ct{Mig91} The numbers came out too far from the experiment and
it was
hard to improve them. Similar attempts with the Gaussian Clebsch
variables
\ct{MW92} also failed to produce the correct numbers. It became clear
to me
that velocity field fluctuates too much to be used as a basic
variable in
turbulence.\footnote{This does not mean that the simple Gaussian
models in
velocity or Clebsch variables could not explain observed turbulence
in finite
systems. We are talking about idealized problem of isotropic
homogeneous
turbulence  with infinite Reynolds number.}

We encounter the same problem in QCD  where the gauge field strongly
fluctuates, and its correlations do not decrease with distance.  The
problem is
not yet solved there, but some useful tools were developed. In
particular, the
Wilson loops (the averages of the ordered exponentials of the
circulation of
the gauge field) are known to be a better field variables.  The
Wilson loop is
expected to be described by some kind of the string theory, though
nobody
managed to prove this.

The dynamical equations for the Wilson loops as  functionals of the
shape of
the loop were derived, and studied for many years\ct{Mig83}.  The QCD
loop equations proved to be very hard to solve, because of the
singularities at self intersections.

In my recent paper\ct{Mig93} I derived similar equations for the
averages
of the exponentials of velocity circulation in (forced) Navier-Stokes
equation. These equations have no singularities at self
intersections, in
addition they are linear, unlike the loop equations of QCD. This
raised
the hopes to find exact solution in terms of the string functional
integrals.

The theory developed below started as the solution the (Euler limit
of
the) loop equations. However, later I found how  to derive it from
the
Liouville equation, whithout unjustified assumptions of the loop
calculus.
This is how I am  presenting this theory here.

\section{Loop functional}

It is generally believed that vorticity is more appropriate  than
velocity
for the description of turbulence. Vorticity is invariant with
respect to
Galilean transformations which shift the space independent part of
velocity
\begin{equation}\label{Ga}
\val(r) \Ra \val(r - u t) + u_{\alpha} ;\; \omega_{\alpha\beta}(r)
\Ra
\omega_{\alpha\beta}(r - u t)
\end{equation}
The correlation functions of velocity field involve the infrared
divergencies coming from this part. Say, in the two- point
correlation
function this would be the energy density
\begin{equation}\label{E0}
\VEV{\val(r)\val(r')} = \frac{2H}{V} - \oh
\VEV{\left(\val(r)-\val(r')\right)^2}
\end{equation}
which formally diverges  as
\begin{equation}\label{Sp}
\frac{H}{V} \propto \int_{1/L}^{\8} d k k^{-\frac{5}{3}} \sim L^{\tt}
\end{equation}
according to Kolmogorov scaling \cite{Kolm41}.

The  infrared divergencies are absent in vorticity correlations. The
corresponding Fourier integral is ultraviolet divergent due to extra
factor of
$ k^2 $, but this is healthy. The physical observables involve the
vorticity
correlations at split points, where the integrals converge.

The complete set of such observables is generated by velocity
circulations for
various loops in the fluid
\begin{equation}\label{Ci}
\Gamma_C[v] = \oint_C d \ral \val(r)
\end{equation}
The constant part of velocity drops here after the integration over
the
closed loop, thus the circulation is Galilean invariant. The loop
gets
translated
\begin{equation}\label{C'}
C \Ra C - u t
\end{equation}
but all the equal time correlations of the circulation stay invariant
in
virtue of translation symmetry.

One can express the circulation in terms of vorticity via  the Stokes
theorem
\begin{equation}\label{St}
\Gamma_C[v] = \sum_{\mu < \nu}\int_S dr_{\mu} \wedge dr_{\nu}
\,\omega_{\mu\nu}(r)
\end{equation}
where $ S $ is an arbitrary surface bounded by $ C $. In particular,
for
infinitesimal loop we would get the local vorticity.
The moments of the circulation
\begin{equation}\label{}
\VEV{\left(\Gamma_C[v]\right)^n} = \int [Dv] P[v]
\left(\Gamma_C[v]\right)^n
\end{equation}
all converge in the ultraviolet as well as in the infrared domain.
The
infrared convergence is guaranteed since these are surface integrals
of
vorticity, and the ultraviolet one is guaranteed since these are line
integrals of velocity.

This nice property suggests to study the distribution of the velocity
circulation
\begin{equation}\label{Pdf}
P_C[\Gamma] = \int [Dv] P[v] \delta\left(\Gamma- \Gamma_C[v]\right)
\end{equation}
It is more convenient to study the Fourier transform
\begin{equation}\label{Psi}
\Y = \int_{-\8}^{\8} d \Gamma \EXP{\i \gamma\Gamma} P_C(\Gamma) =
\int [Dv] P[v] \EXP{\i \gamma \Gamma_C[v] }
\end{equation}
We expect these functionals to exist in the turbulent
limit unlike the distribution $ P[v] $.

The dynamical equation for these functionals (the loop equation) was
derived
in my previous work\ct{Mig93}.
The time derivative of the circulation reads
\be
\dt \Gamma_C[v] = \C \ral \omega_{\alpha\beta}(r) \vbe(r)
\label{dG}
\ee
All the nonlocal terms reduced to the closed loop integrals of the
total
derivatives and vanished.
Being averaged with appropriate measure, the remaining terms in time
derivative must vanish according to the Liouville equation
\be
\VEV{ \dt \Gamma_C[v] \, \EXP{\i g  \Gamma_C[v]}} = 0
\ee

Note that this is {\em not} the Kelvin theorem of
conservation of the circulation. The circulation is conserved in a
Lagrange sense, at purely kinematical level
\be
\dd{t} \Gamma_{C}[v] = \dt \Gamma_{C}[v] +
\oint d \theta
\fbyf{\Gamma_{C}[v]}{C_{\beta}(\theta)}\vbe\left(C(\theta)\right) =0
\ee
In other words, the Euler derivative \rf{dG}  is exactly compensated
by the
shift of every point at the loop by local velocity.
Now, according to the Liouville equation, {\em each} of these equal
terms must vanish in average.

The formal derivation goes as follows
\bea
\int [Dv] P[v] \EXP{\i \gamma \Gamma_C[v] } \PBR{\i
\gamma\Gamma_C[v],\,
H} \br
=
\int [Dv] P[v] \PBR{\EXP{\i \gamma \Gamma_C[v] }, H}  \br
=
-\int [Dv] \PBR{P[v],H} \EXP{\i \gamma \Gamma_C[v] } = 0
\eea
and the physics is obvious: the average of any time derivative with
time
independent weight must vanish.

Let us interpret in these terms the solution for $ \Y $ found in the
previous paper \ct{Mig93}
\be
\Y = f(\Sigma_{\alpha\beta});\; \Sigma_{\alpha\beta} = \C \ral \rbe
\ee
In virtue of linearity of the Liouville equation it suffices to check
the
Fourier transform
\be
\EXP{\i \gamma R_{\alpha\beta} \Sigma_{\alpha\beta}}
\ee
In terms of the velocity field this corresponds to global rotation
\be
\val = R_{\alpha\beta} \rbe ;\; \omega_{\alpha\beta}  = 2
R_{\alpha\beta}
\ee
The corresponding \PB reads
\be
\PBR{\Gamma_C[v], H} = \C \ral \omega_{\alpha\beta} \vbe = 4
\Sigma_{\alpha\gamma} R_{\alpha\beta} R_{\beta\gamma} =0
\ee
The sum over tensor indexes vanished by symmetry.

This solution is always present in fluid mechanics due to the
conservation of the angular momentum. Unfortunately, it has nothing
to do
with turbulence, contrary to the hopes expressed in my previous work.

Let us also note, that the Gibbs solution does not apply here. One
could
formally compute the loop functional for the Gibbs distribution, but
the
result is singular\footnote{This is yet another advantage of the loop
functional:
the Gibbs solution for the loop functional
does not exist, which forces us to look for alternative invariant
distributions.}
\be
\Y = \EXP{ - \frac{\gamma^2}{2\beta} \C \ral \C \ral' \delta^3(r-r')}
\ee
Should we cut off the wavevectors at $ k \sim \frac{1}{r_0} $ this
would
become
\be
\Y \approx \EXP{ - \frac{\gamma^2}{\beta r_0} \oint_C |dr|}
\ee

This is so called perimeter law, characteristic to the local vector
fields.
Clearly this is not the case in turbulence, as velocity field is
highly
nonlocal. Also, the odd correlations of velocity, such as the triple
correlation, which are present due to the time irreversibility,
would, in
general make the loop functional complex.

\section{Vortex dynamics}

Let us study the dynamics of the vortex structures from the
Hamiltonian
point of view. We shall assume that vorticity is not spread all over
the
space but rather occupies some fraction of it. It is  concentrated in
some number of cells $ D_i $ of various topology moving in their own
velocity field.

Clearly, this picture is an idealization. In the real fluid, with
finite
viscosity, there will always be some background vorticity between
cells. In
this case, the cells could be defined as the domains with vorticity
above this
background. The reason we are doing this is obvious: we would like to
work with
the Euler equation with its symmetries.

We shall use two types of tensor and vector indexes. The Euler (fixed
space)
tensors will be denoted by Greek subscripts such as $ \ral $. The
Lagrange
tensors (moving with fluid) will be denoted by latin subscripts such
as $
\rho_a $. The  field $ X_{\alpha}(\rho) $ describes the instant
position of the
point with initial coordinates $ \rho_a $. The transformation matrix
from the
Lagrange to Euler frame is given by $ \d_a X_{\alpha}(\rho) $.

The contribution of each cell $ D $ to the net velocity field can be
written as follows
\be
\val(r) = -e_{\alpha\beta\gamma}\d_{\beta}
\int_D  d^3 \rho  \frac{\Omega_{\gamma}(\rho)}{4\pi|r-X(\rho)|}
\ee
where
\be
\Omega_{\gamma}(\rho) = \Omega^a(\rho)\d_a X_{\gamma}(\rho)
\ee
is the vorticity vector in the Euler frame.
The vorticity vector $ \Omega^a(\rho) $ in the  Lagrange frame is
conserved
\be
\dt \Omega^a(\rho) = 0
\ee

The  physical vorticity tensor $ \omega_{\alpha\beta} = \dal
\vbe-\dbe\val $
inside the cell can be readily computed from velocity integral. The
gradients produce the $ \delta $ function so that we get
\be
\omega_{\alpha\beta}(X(\rho)) =
\left(\JAC{X_1,X_2,X_3}\right)^{-1}\,e_{\alpha\beta\gamma}
\Omega_{\gamma}(\rho)
\ee
or in Euler frame, inverting $ \rho= X^{-1}(r)  $
\be
\omega_{\alpha\beta}(r) = \frac{1}{6}\,e_{\lambda\mu\nu}\,e_{abc}
\d_{\lambda} \rho^a \d_{\mu}\rho^b \d_{\nu} \rho^c \,
e_{\alpha\beta\gamma}\,
\Omega^f(\rho)\d_f X_{\gamma} = e_{abc} \dal\rho^a(r) \dbe \rho^b(r)
\,
\Omega^c(\rho(r))
\ee
The inverse matrix $ \dal \rho^a = \left(\d_a
X_{\alpha}\right)^{-1} $ relates the Euler  indexes to the Lagrange
ones,
as it should.
These relations between the Euler and Lagrange vorticity are
equivalent to
the conservation of the vorticity 2-form
\be
\Omega = \sum_{\alpha<\beta}\omega_{\alpha\beta} (X)
dX_{\alpha}\wedge
dX_{\beta} =
\sum_{a<b}\Omega_{ab} (\rho)d\rho^a \wedge d\rho^b
;\;\Omega_{ab} = e_{abc} \Omega^c
\ee
which is the Kelvin theorem of the conservation of velocity
circulation
around arbitrary fluid loop.

The field $ X_{\alpha}(\rho) $  moves with the flow (the Helmholtz
equation)
\be
\dt X_{\alpha}(\rho) = \val(X(\rho))
\ee
In Appendix B we derive the Euler equation from the Helmholtz
equation
and study the general properties of the former. We  show that this is
equivalent to the Hamiltonian dynamics with the (degenerate) Poisson
brackets
\be
\PBR{X_{\alpha}(\rho), X_{\beta}(\rho')} = -\delta^3(\rho-\rho')
e_{\alpha\beta\gamma}
\frac{\Omega_{\gamma}(\rho)}{\Omega_{\mu}^2(\rho)}
\ee
The Hamiltonian is given by the same fluid energy, with the velocity
understood
as functional of $ X(.) $ The degenaracy  of the Poisson brackets
reflects the
fact that there are only two independent degrees of freedom at each
point.
This leads to the gauge symmetry which we discuss below.

The Hamiltonian variation reads
\be
\frac{\delta H}{\delta X_{\alpha}(\rho)} =
e_{\alpha\beta\gamma} \vbe\left(X(\rho)\right)\Omega_{\gamma}(\rho)
\label{dH}
\ee
This variation is orthogonal to velocity, which provides the energy
conservation. It is also orthogonal to vorticity vector which leads
to the
gauge invariance. The gauge transformations
\be
\delta X_{\alpha}(\rho) = f(\rho) \Omega_{\alpha}(\rho)
\ee
leave the Hamiltonian invariant. These transformations reparametrize
the
coordinates
\be
\delta\rho^a = f(\rho) \Omega^a(\rho)
\ee
The vorticity density transforms as follows
\be
\delta \Omega^a = - \Omega^a \d_b ( f \Omega^b ) + f \Omega^b \d_b
\Omega^a + \Omega^b \d_b (f \Omega^a) = 2 f \Omega^b \d_b \Omega^a
\ee
The first term comes from the volume transformation, the second one -
from the argument transformation and the third one - from the vector
index transformation of $ \Omega $  so that
\be
d^3 \rho \,\Omega^a \d_a = \mbox{inv}
\ee
The identity
\be
\d_a \Omega^a = 0
\ee
was taken into account. We observe that these gauge transformations
leave
invariant the whole velocity field, not just the Hamiltonian.

The vorticity 2-form simplifies in the Clebsch variables
\be
\Omega^a = e^{abc} \d_b\phi_1(\rho) \d_c\phi_2(\rho)
\label{CL}
\ee
\be
\Omega =  d \phi_1 \wedge d\phi_2  = \mbox{inv}
\ee
The Clebsch variables provide the bridge between the Lagrange and the
Euler
dynamics. By construction they are conserved, as they parametrize the
conserved
vorticity. The Euler Clebsch fields $ \Phi_i(r) $ can be introduced
by solving
the equation $ r = X(\rho) $
\be
\Phi_i(r) = \phi_i\left(X^{-1}(r)\right)
\ee

However, unlike the vorticity field, the Clebsch variables cannot be
defined
globally in the whole space. The inverse map $ \rho = X^{-1}(r) $ is
defined
separately for each cell, therefore one cannot write $ \val = \Phi_1
\dal
\Phi_2 + \dal \Phi_3 $ everywhere in space. Rather one should add the
contributions from all cells to the Poisson integral, as we did
before.

This explains the notorious helicity paradox. The conserved helicity
integral
\be
 {\cal H} =  \int d^3 \rho\, v_c(\rho)\, \Omega^c(\rho)
 \ee
where
\be
v_a(\rho) = \phi_1 \d_a \phi_2 + \d_a \phi_3; \d^2 \phi_3 = -
\d_a\left( \phi_1
\d_a \phi_2\right)
\ee
is the initial velocity field (to be distinguished from the physical
velocity
field $\val(r) $ which cannot be paramatrized globally by the Clebsch
variables).

The helicity integral for the finite cell could be finite. It can be
written in
invariant terms of the vorticity forms
\be
{\cal H_D} = \int_D d^3 \rho e^{abc} \d_a \phi_1 \d_b \phi_2 \d_c
\phi_3 =
\int_D d\phi_1\wedge d\phi_2 \wedge d\phi_3 = \int_{\d D}
\phi_3\,d\phi_1
\wedge d\phi_2
\ee
from which representation it is clear that it is conserved
\be
{\cal H_D} = \int_{\d D}\Phi_3\, d\Phi_1 \wedge d\Phi_2
\ee

Our gauge transformation leave the Clebsch field invariant
\be
\delta \phi_i(\rho) = f(\rho) \Omega^a(\rho)\d_a \phi_i(\rho) = 0
\ee
The velocity integral in Clebsch variables reduces to the 3-form
\be
\val(r) = -e_{\alpha\beta\gamma}\d_{\beta}
\int_D   \frac{dX_{\gamma}\wedge d\phi_1\wedge d\phi_2 }{4\pi|r-X|}
\ee

This gauge invariance is less than the full diffeomorphism group
which
involves arbitrary function for each component of $ \rho$.
This is a subtle point. The field $ \Omega^a(\rho) $ has no dynamics,
it
is conserved. However, the initial values of $ \Omega^a $ can be
defined only
modulo diffeomorphisms, as the physical observables are parametric
invariant. So, we could as well average over reparametrizations of
these
initial values, which would make the parametric invariance complete.

Another subtlety. The equations of motion, which literally follow
from
above Poisson brackets describe the motion in the direction
orthogonal to  the
gauge transformations, namely
\be
\dt X_{\alpha}(\rho) = \left(\delta_{\alpha\beta} -
\frac{\Omega_{\alpha}(\rho)\Omega_{\beta}(\rho)}{\Omega_{\mu}^2(\rho)
}\right)
\vbe\left( X(\rho)\right) = \val(X(\rho))  + f(\rho)
\Omega_{\alpha}(\rho)
\ee
The difference is unobservable, due to gauge  invariance.  We could
have
defined the Helmholtz equation this way from the very beginning.
The conventional Helmholtz  dynamics represents so called generalized
Hamiltonian dynamics, which cannot be described in terms of the
Poisson
brackets. The formula \rf{dH} cannot be solved for the velocity field
because the matrix
\be
\Omega_{\alpha\beta}(\rho) =
e_{\alpha\beta\gamma}\Omega_{\gamma}(\rho)
\ee
cannot be inverted (there is the zero mode $ \Omega_{\beta}(\rho) $).
The physical meaning is the gauge invariance which allows us to
perform the
gauge transformations in addition to the Lagrange motion of the fluid
element.

Formally, the inversion of the $ \Omega-$matrix can be performed in a
subspace
which is orthogonal to the zero mode. The inverse matrix
$\Omega^{\beta\gamma}$
in this subspace satisfies the equation
\be
\Omega_{\alpha\beta} \Omega^{\beta\gamma} = \delta_{\alpha\gamma} -
\frac{\Omega_{\alpha}\Omega_{\gamma}}{\Omega_{\mu}^2}
\ee
which has the unique solution
\be
\Omega^{\beta\gamma} = - e_{\alpha\beta\gamma}
\frac{\Omega_{\alpha}}{\Omega_{\mu}^2}
\ee
This solution leads to our Poisson brackets.

Now we see how the correct number of degrees of freedom is restored.
The
Hamiltonian vortex dynamics locally has only two degrees of freedom,
those
orthogonal to the gauge transformations. We could have obtained the
same \PB by
canonical transformations in the vorticity form from the Clebsch
variables to
the $ X-$variables. This canonical transformation describes the
surface in the
$ X-$space. The vorticity form at this surface can be treated as a
degenerate
form in the 3-dimensional space.

One may readily check  the conservation of the volume element of the
cell
\be
\dt \JAC{X_1,X_2,X_3} = \JAC{X_1,X_2,X_3}\,\d_{\alpha} \val(X) = 0
\ee
Using the formulas for the variations of the velocity field derived
in
Appendix B  one may prove the stronger statement of the phase volume
element conservation (Liouville theorem)
\be
\fbyf{\val\left(X(\rho)\right)}{X_{\alpha}(\rho)} = 0
\ee
\be
(DX) = \prod_{\rho} d^3X(\rho)= \mbox{ const}
\ee

\section{Vortex statistics}

Let us recall the foundation of statistical mechanics. The Gibbs
distribution can be formally derived from the Liouville equation plus
extra requirement of multiplicativity. In general, any additive
conserved
functional $ E $ could serve as energy in the Gibbs distribution.

The physical mechanism is the energy exchange between the small
subsystem
under study and the rest of the system (thermostat). The conditional
probability for a subsystem is obtained from the microcanonical
distribution $ d\Gamma' d \Gamma\delta(E' + E - E_0) $ for the whole
system by
integrating out the configurations $ d\Gamma' $ of the thermostat.

The corresponding phase space volume $ e^{S(E')} = \int
d\Gamma'\delta(E' + E -
E_0)   $ of the thermostat depends upon its energy $ E' = E_0 -  E $
where the
contribution $ - E $ from the subsystem represents the small
correction.
Expanding $
S(E_0-E) = S(E_0) - \beta E $ we arrive at the Gibbs distribution.

In case of the vortex statistics we may try the same line of
arguments. The
important addition to the general Gibbs statistics is the parametric
invariance. Reparametrizations, or gauge transformations, are part of
the
dynamics, as have seen above.\footnote{This makes so hard the
numerical
simulation of the Lagrange motion. The significant part of notorious
instability of the Lagrange  dynamics is the reparametrization of the
volume
inside the vortex cell.}
So, the Gibbs distribution should be both gauge invariant and
conserved.

The net volume of the set of vortex cells $ V = \sum_{i} V(D_i) $
where
\be
 V(D) = \int_{D}  d^3 \rho \JAC{X_1,X_2,X_3} =
\int_{D}  dX_1\wedge dX_2 \wedge d X_3
\ee
is the simplest term in effective energy of the Gibbs distribution.
It is gauge invariant, additive and positive definite.It is bounded
from above
by the volume of the system, so that there could be no infrared
divergencies.\footnote{The Hamiltonian does not exist in the
turbulent flow
because the energy spectrum diverges at small wavevectors. The net
Hamiltonian
grows faster than the volume of the system which is unacceptable for
the Gibbs
distribution.}

The mechanism leading to the thermal equilibrium is
quite transparent here. The volume of a little cell surrounded by the
large amount of other cells, would fluctuate due to exchange with
neighbors.\footnote{The volume, as well as any other local functional
of the
cell does not change at splitting/joining, therefore these processes
go with
significant probability. For the nonlocal functionals, such as the
hamiltonian,
there are long range interactions, which makes the exchange process
less
probable.}These are the viscous effects, in the same way as the
energy
exchange mechanisms in the ideal gas were the effects of interaction.
The
relaxation time is inversely proportional to the strength of
interaction
(viscosity in our case).

We must take these effects into account in kinetics, but the
resulting
statistical distribution involves only the energy of the ideal
system. This
was the most impressive part of the achievement of Gibbs and
Boltzman. They
found the shortcut from mechanics to statistics, avoiding the
kinetics.
All the interactions are hidden in the temperature and chemical
potentials.

Mechanically, the cells avoid each other as well as themselves and
preserve their topology  but the implicit viscous interactions would
lead
to fluctuations. Even if we start from one spherical cell
it would inevitably touch itself in course of the time evolution. At
the
vicinity of the touching point the viscous effects show up, which
break
the topological conservation laws of the Euler dynamics. The result
could
be a handle, or  the splitting into two cells. After long  evolution
we
would end up with the ensemble of cells $ D_i $ with various number $
h_i $
of handles.

The related subject is the vorticity vector field $ \Omega^a(\rho) $
inside the cells. In the Euler dynamics it is conserved, but the
viscosity-generated interaction would lead to fluctuations. The
invariant
measure is
\be
(D\Omega) = \prod_{\rho} d^3\Omega(\rho) \delta[\d_a \Omega^a(\rho)]
\ee
In terms of the Clebsch variables \rf{CL} the measure is simply
\be
(D\phi) = \prod_{\rho}d^2\phi(\rho)
\ee
as these are canonical hamiltonian variables. As discussed above,
there are no
global Clebsch variables in Euler sense. These Clebsch variables are
defined
separately in each cell and the net velocity field is a sum of
contributions
from all cells rather than the single Clebsch-parametrized expression
$ \Phi_1
\dal \Phi_2 + \dal \Phi_3 $.

What could be an effective energy for the vorticity? The helicity
integral is
excluded as a pseudoscalar, besides, it is nonlocal, like the
Hamiltonian.
We insist on parametric invariance and locality in a sense that the
cell
splitting and joining do not change this energy. The Clebsch
variables are
defined modulo additive constants, they could be multivalued in
complex
topology, therefore we have to use the vorticity field itself. The
generic $
\Omega-$invariant in $ d $ dimensions is
\be
 \int_{D}d^d \rho  \sqrt{\mbox{det } \Omega_{ab}}
\ee
In {\em even} dimensions $ d = 2k$ this invariant reduces to the
Pfaffian
\be
\int_D d^{2k} \rho \sqrt{\mbox{det } \Omega_{ab}} =
\frac{1}{k!}\int_D \Omega\wedge\Omega\dots \wedge \Omega =
\int_{D}\frac{\Omega^k}{k!}
\ee
In two dimensions it would be simply the net vorticity of the cell
\be
\int_D d^2 \rho  \Omega_{12}
\ee
However, in odd dimensions it vanishes so that there is no
$\Omega-$invariant.
We see that  there is a significant difference in the vortex
statistics in even
and odd dimensions.

Another interesting comment. For odd $ k = \frac{d}{2} $  the $
\Omega-$invariant is an odd functional of $ \Omega$ which breaks time
reversal
invariance. The vorticity  stays invariant under space reflection but
changes
sign at time reversal. This simple local mechanism of the time
irreversibility
is present only at $ d = 4 k + 2, k=0,1,\dots $. In three dimensions
we live in
it is absent.

Let us turn to the boundary terms. The boundary of the cell $ S = \d
D $  is
described by certain parametric equation
\be
S: \rho_a = R_a\left(\xi_1,\xi_2\right)
\ee
Clearly, in our case this is a self- avoiding surface.
Here we could add the following local surface terms to the energy
\be
E_{\phi} = \sum_i\int_{\d D_i} d^2\xi  \left(a \sqrt{g}+ b \sqrt{g}
g^{ij}
\d_j\phi_k
\d_j \phi_k \right)
\ee
where
\be
g_{ij}(\xi) =  \d_i R_a(\xi)\d_j R_a(\xi)
\ee
is the induced metric.

There is also a topological term for each non-contractible loop $ L $
of $
\d D $
\be
E_{\Theta} = \Theta\,\int_{L} \phi_1 d \phi_2
\ee
This is simply the velocity circulation around such loop. Assuming
continuity
(i.e. vanishing) of vorticity at the boundary of the cell, this
circulation can
be also written as the integral in the external space. Then is it
obvious that
this integral does not depend upon the shape of the loop $ L $, it is
given by
the invariant vorticity flux through the cross section $ \Sigma $ of
the
corresponding  handle
\be
E_{\Theta} = \Theta\,\int_{\Sigma} \Omega;\; L = \d \Sigma
\ee
This term breaks the time reversal! This is the only possible source
of the
irreversibility in this theory in three dimensions.

The following observation leads to crucial simplifications. The only
term which
depends upon the Lagrange field $ X_{\alpha}(\rho) $ is the volume
term, which
in fact is the functional of the bounding surface $ X(\d D) $
\be
V = \int_D d X_1 \wedge d X_2 \wedge d X_3 = \int_{\d D} X_3 \, d X_1
\wedge d
X_2
\ee
The rest of the terms in the effective energy also depending only
upon the
boundary, this lowers the dimension of our effective field theory. We
are
dealing with the theory of self-avoiding random surfaces rather that
the 3-d
Lagrange dynamics. With some modifications the methods of the string
theory can
be applied to this problem.

The invariant distance in the $ X(\xi) $ functional space is
\be
||\delta X ||^2 = \int_{\d D} d^2 \xi  \sqrt{g}  \left(\delta
X_{\alpha}(\xi)\right)^2
\ee
where $ g $ as usual stands for the determinant of the metric tensor.
One may check that the corresponding volume element
\be
(DX) = \prod_{\xi \in \d D} \delta X
\ee
is conserved in the Euler-Helmholtz dynamics as well as the complete
volume
element. The key point in this extension of the Liouville theorem is
the
observation that the matrix trace of the functional derivative
vanishes
\be
\fbyf{\val(X(\rho))}{X_{\alpha}(\rho')} = 0
\ee
for arbitrary $ \rho, \rho'$, including the boundary points. The
metric tensor
$ g_{ij} $ does not introduce any complications, as it is  $ X $
independent.

This metric is the motion invariant in the vortex dynamics. In the
statistics, according to the general philosophy these invariants
become
variables. We see that the field $ R_a(\xi) $ enter only via the
induced
metric, which allows us to introduce the latter as a collective field
variable.

One has to introduce the functional space of all metric tensors  with
the
Polyakov distance \ct{Pol}
\be
||\delta g ||^2 = \int_S d^2 \xi  \sqrt{g}  \left(\delta g_{ij}
\delta
g_{kl}\right) \left(A g^{ij} g^{kl} + B g^{ik} g^{jl}\right)
\ee

The parametric invariance can be most conveniently fixed by the
conformal gauge
\be
g_{ij}(\xi)  = \hat{g}_{ij}(\xi) e^{ \alpha\varphi(\xi)}
\ee
where $ \hat{g} $ is some background metric parametrizing the surface
with given topology. Unlike the internal metric, the background
metric
does not fluctuate. We have the freedom to choose any parametrization
of
the background metric.

The effective energy which emerges after all computations of the
functional jacobians associated with the gauge fixing reads\footnote{
We
cut some angles here. This effective measure was obtained \ct{DDK}
with
some extra locality assumptions in addition to the honest
calculations.
These assumptions were never rigorously proven, but they are known to
work. All the results which were obtained with this measure coincide
with those
obtained by the mathematically justified method of dynamical
triangulations and matrix models.}
\be
E_{\varphi} = \frac{1}{4\pi} \int d^2 \xi
\sqrt{\hat{g}}\left(\oh \hat{g}^{ij}\d_i \varphi \d_j \varphi - Q
\hat{R} \varphi + \mu e^{\alpha\varphi} \right)
\ee
where $ \hat{R} $ is the scalar curvature in the background metric.

The parameters $ Q, \alpha $ should be found from the
self-consistency
requirements. In case of the ordinary string theory in $ d $
dimensional
space the requirement of cancellation of conformal anomalies yields
\cite{DDK}
\be
\alpha = \frac{\sqrt{1-d} - \sqrt{25-d}}{2 \sqrt{3}};\;
Q = \sqrt{\frac{25-d}{3}}
\ee
In three dimensions $ \alpha $ is a complex number, which is fatal
for the string theory. Fortunately this formula does not apply to
turbulence, because the dynamics of the $ X $ field is completely
different here. Later we speculate about the values of these
parameters.

To summarize, the total phase space volume element of our string
theory
\be
d \Gamma = \prod_{\mbox{cells}} (DX)(D\phi)(D\varphi)
\ee
and the total effective energy
\be
E = \sum_{\mbox{cells}} \left(V  + E_{\phi} + E_{\varphi}
+\sum_{\mbox{loops}}
E_{\Theta}  \right)
\ee
The grand partition  function
\be
Z = \sum_{N,h_i} \EXP{ -\mu N - \lambda \sum_{i=1}^N h_i}\int
d\Gamma
\EXP{- \beta E}
\ee
The interaction between cells comes from the excluded volume effect.

The vorticity correlations are generated by the loop
functional\ct{Mig93}
\be
\Psi_C(\gamma) = \VEV{\EXP{\i \Gamma_C[v]}}
\ee
where
\be
\Gamma_C[v] = \C \ral \val(r)
\ee
is the velocity circulation.
Our solution for the loop functional reads
\be
\Psi_C(\gamma) =  Z^{-1} \sum_{N,h_i} \EXP{ -\mu N - \lambda
\sum_{i=1}^N
h_i}\int' d\Gamma \EXP{- \beta  E +\i \gamma \Gamma_C[v] }
\ee
where $ \int' $ implies that the cells also avoid the loop $ C $.

There are now various topological sectors. In the trivial sector, the
loop can be contracted to a point without crossing the cells.
Clearly,
circulation vanishes in this sector. In the nontrivial sectors, there
is
one or more handles encircled by the loop, so that the circulation is
finite. Here is an example of such topology

\let\picnaturalsize=N
\def\picsize{3.0in}
\def\picfilename{2tori.eps}
\ifx\nopictures Y\else{\ifx\epsfloaded Y\else\input epsf \fi
\let\epsfloaded=Y
\centerline{\ifx\picnaturalsize N\epsfxsize \picsize\fi
\epsfbox{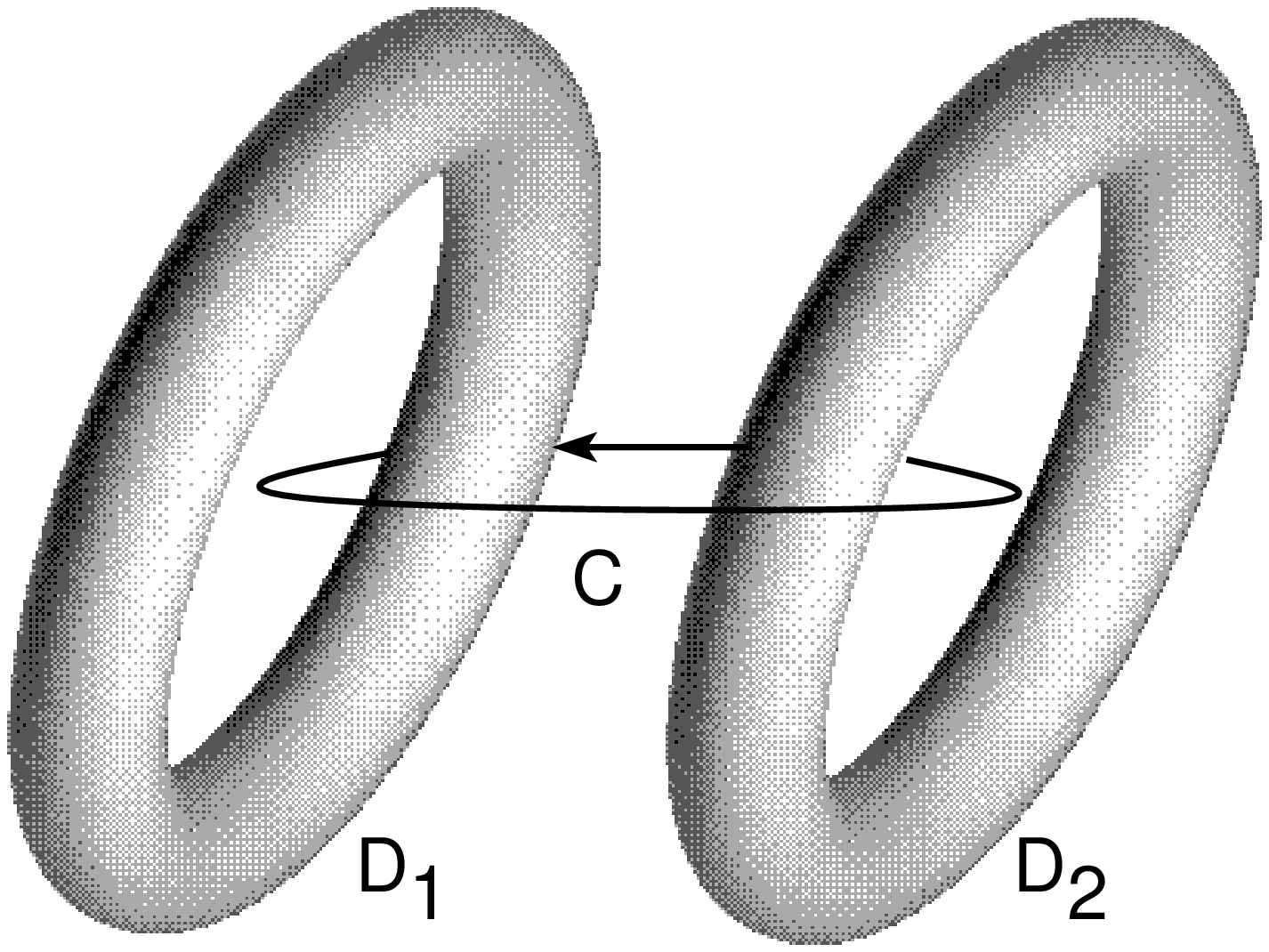}}}\fi

The circulation can  be reduced to the Stokes integral of the
vorticity 2-form
over the surface $ S_C $ encircled by the loop $ C $. Only the parts
$ s_i =S_C
\cap D_i $ passing through the cells contribute. In each such part
the
vorticity
2-form can be transformed to the Clebsch coordinates which reduces it
to the
sum of the topological terms
\be
\Gamma_C = \sum_{i} \int_{s_i} \Omega =
\sum_{i} \int_{s_i} d\phi_1 \wedge d \phi_2  = \sum_{i} \int_{\d s_i}
\phi_1 d
\phi_2
\ee
In presence of the same terms in the effective energy the loop
functional would
be complex, as the positive and negative circulations would be
weighted
differently. This is manifestation of the time irreversibility.

The interesting thing is that the interaction between the $ X $ field
and the
rest of field variables originates in the topological restrictions.
The
circulation reduces to the net flux from the handles encircled by the
loop in
physical space (this is a restriction on the $ X $ field). Also, the
requirement that the loop is avoided by cells imposes another
restriction on
the $ X $ field. There are no explicit interaction terms in the
energy. The
whole dependence of the loop functional on the shape of the loop $ C
$ comes
from the excluded volume effect. The implications of this remarkable
property
are yet to be understood.

Let us now check the Liouville equation. The time derivative of the
circulation
reads (see\rf{dG})
\be
\dt \Gamma_C[v] = \C \ral \omega_{\alpha\beta}(r) \vbe(r)
\label{dG'}
\ee
But the vorticity  vanishes at the loop, because the cells avoid it!
Therefore
\be
\VEV{ \dt \Gamma_C[v] \, \EXP{\i \gamma  \Gamma_C[v]}} = 0
\ee
which is the Liouville equation.

For the readers of the previous paper\ct{Mig93} let us briefly
discuss
the loop equation. The essence of the loop equation is the
representation
of the vorticity as the area derivative acting on the loop functional
\be
\hat{\omega}_{\alpha\beta}(r) = - \frac{\i}{\gamma}\,
\ff{\sigma_{\alpha\beta}(r)}
\ee
The formal definition of the area derivative in terms of the ordinary
functional derivatives was discussed before \ct{Mig83,Mig93}. The
geometric meaning is simple: add the little loop to the original one
and
find the term linear in the area $ \delta \sigma_{\alpha\beta} $
enclosed
by this little loop. The parametric invariant functionals like this
one
can always be regularized so that the variation would start from $
\delta
\sigma_{\alpha\beta} $. The area derivatives of the length and the
minimal
area inside the loop were derived in \ct{Mig83}.

The velocity operator is related to vorticity by the Poisson integral
\be
\hat{v}_{\alpha}(r) =-
\dbe \int d^3 R \frac{\hat{\omega}_{\alpha\beta}(R)}{4\pi|r-R|}
\ee
The geometric meaning is as follows. The vorticity operator adds the
little loop $ \delta C_{R,R} $ at the point $ R $ off the original
loop $ C $.
By adding the couple of straight line integrals
\be
W(r,R) = \EXP{\i \gamma  \int_{L_{r,R}} d \ral \val(r)} = W^{-1}(R,r)
\ee
we reduce this to one loop of the singular shape
\bigskip
\let\picnaturalsize=N
\def\picsize{3.0in}
\def\picfilename{Velocity.eps}
\ifx\nopictures Y\else{\ifx\epsfloaded Y\else\input epsf \fi
\let\epsfloaded=Y
\centerline{\ifx\picnaturalsize N\epsfxsize \picsize\fi
\epsfbox{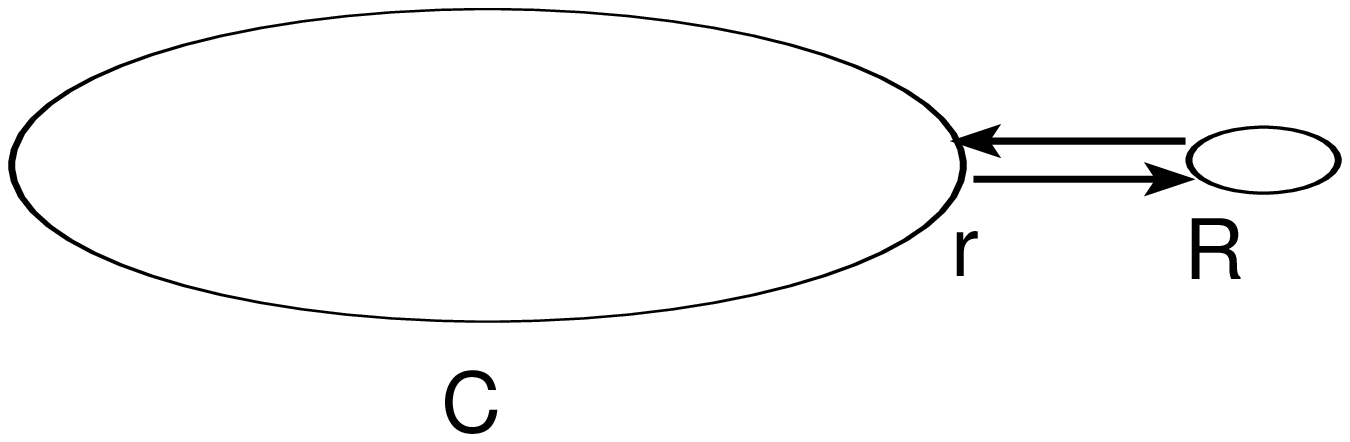}}}\fi
\be
\tilde{C} =\left\{C_{r,r},L_{r,R},\delta C_{R,R}, L_{R,r}\right\}
\ee
\be
\Gamma_C  + \Gamma_{\delta C_{R,R} } = \Gamma_{\tilde{C}}
\ee
which can be obtained from original one as certain
variation\ct{Mig93}.
Then the loop equation is simply
\be
\C \ral \hat{\omega}_{\alpha\beta}(r) \hat{v}_{\beta}(r) \Y = 0
\ee
It was shown in \ct{Mig93} that these two operators commute as they
should.

In our case this equation is satisfied in a trivial way.
Regardless the subtleties in the definition of $ \hat{\omega},\hat{v}
$,
as long as these operators act on the vortex sheet circulation as
they
should, they insert the vorticity and velocity at the loop.
The rest of the argument is the same as in the Liouville equation.

Note that this trick would not solve  the Hopf equation for the
velocity
generating functional
\be
H[J] = \VEV{\EXP{\int d^3 r J_{\alpha}(r) \val(r)}}
\ee
The Hopf equation would require
\be
\VEV{ \int d^3 r'
T_{\alpha\beta}\left(r-r'\right)  \omega_{\beta\gamma}(r') \vga(r')
\EXP{\int d^3 r J_{\alpha}(r) \val(r)}} = 0
\ee
which we cannot satisfy since the volume integral $ \int d^3 r' $
would
inevitably overlap with cells where vorticity is present.

Apparently we found invariant probability distribution for vorticity
but
not for the velocity. The velocity distribution may not exist in the
infinite system, due to the infrared divergencies. In practice this
would
mean that the velocity distribution would depend of the details of
the
large scale energy pumping, but vorticity distribution would be
universal.

The correlation functions of vorticity field can be obtained from
the
multiloop functionals by contracting loops to points. In case when
the
loops encircle one or more handles there would be nonvanishing
correlation function.  Here is an example of the topology with the
two point
correlation which also has nontrivial helicity because of the knotted
handle.
\let\picnaturalsize=N
\def\picsize{3.0in}
\def\picfilename{knot.eps}
\ifx\nopictures Y\else{\ifx\epsfloaded Y\else\input epsf \fi
\let\epsfloaded=Y
\centerline{\ifx\picnaturalsize N\epsfxsize \picsize\fi
\epsfbox{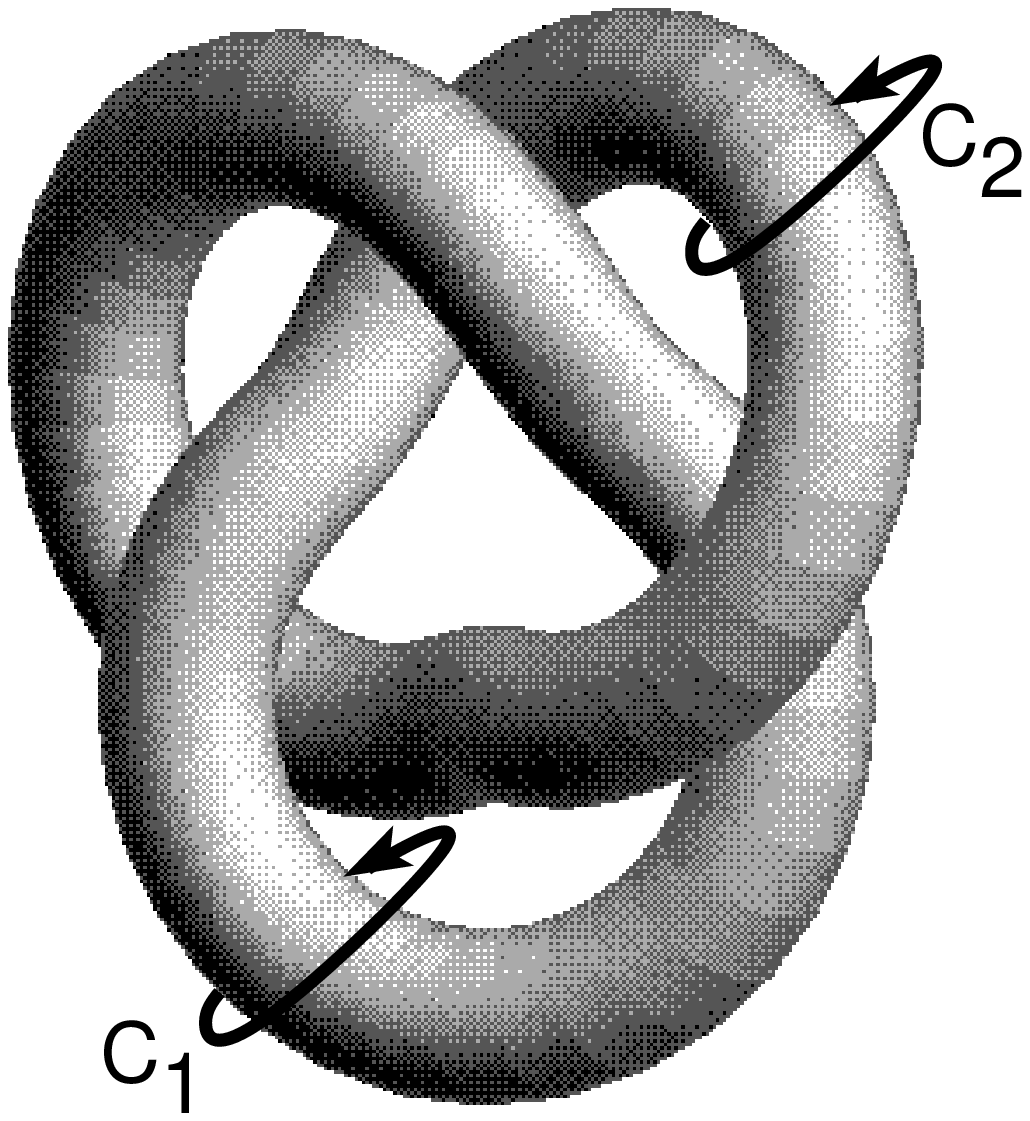}}}\fi

\section{Discussion}

So far we do not have much to tell to the engineers. Even if this
statistics of the vortex structures  will  be confirmed by further
study,
we shall face the formidable task of computing the correlation
functions. Let us speculate what could come out of these
computations.

The qualitative picture of intermittent distribution of vorticity
will be
the same as in the multifractal models\ct{SBconf}. In fact these
models
inspired our study to some extent. Our basic idea is that the part $
V $ of
the volume occupied by vorticity fluctuate. In numerical and real
experiments\ct{SBconf} the high vorticity structures were clearly
seen. The
cells take the shape of long sausages rather than spheres, which does
not
contradict our general philosophy but still lacks an explanation.

Let us try to estimate the intermittency effects in the
vortex cells statistics. Let us contract the loop to a point $ r $
around some
handle. What we get in the limit can be expressed in terms of  the
{\em vertex
operator} of the string theory
\be
\Gamma_C \propto \int_{S} d^2 \xi_0 \sqrt{g}\int_{S} d^2 \eta_0
\sqrt{g}\delta^3(X(\xi) -r) \delta^3(X(\eta) - r');\; r' \ra r
\ee
The points $ \xi_0 $ and $ \eta_0 $ are mapped to the same point $ r'
\ra r $
in physical space: this is the handle strangled by the loop. The
important
detail here is the factor $ \sqrt{g} = e^{\alpha \varphi} $
corresponding to
the metric tensor at the surface.

The properties of such metric tensors were studied in the string
theory\ct{Pol,DDK}. The moments of $ \Gamma_C $ would behave as
\be
\VEV{\Gamma_C^n} \propto r_0^{-\Delta(n)}; \Delta(n) = \oh n\alpha
(n\alpha+Q)
\ee
where $ r_0 $ is the short-distance cutoff.
The parameters $ \alpha,Q $ are to be found from the selfconsistency
conditions.

In case of the turbulence theory we know that the third moment has no
anomalous dimension, due to the Kolmogorov's $ \frac{4}{5} $ law.
This
implies that
\be
Q = -3 \alpha
\ee
after which we exactly reproduce the anomalous dimensions of
so called Kolmogorov-Obukhov intermittency model.

They assumed in 1961 the log-normal distribution of the
energy dissipation rate as a modification of the Kolmogorov scaling.
Various multifractal models generalizing this distribution,
and the corresponding experimental and numerical data are discussed
in proceedings of the 1991 conference in Santa Barbara \ct{SBconf}.

The physics of the fractal dimensions in our theory is almost the
same as in
the Kolmogorov-Obukhov model. The local energy dissipation rate at
the edge can
be estimated as $ \Gamma_C^3 \propto e^{ 3 \alpha \varphi} $. This
quantity is,
indeed, an exponential of the gaussian fluctuating variable. The
variance is
proportional to $ \log r $ since it is the two dimensional field with
the
logarithmic propagator (how could they have guessed that!).

Strictly speaking, the dimensions of powers of $ \Gamma_C $ were
never measured. People considered the moments of velocity differences
instead. In principle, the potential part of $ v $ which is present
in
these moments may change the trajectory, so we must be cautious.

The point is, the potential part has completely different origin in
our
theory. It comes from the nonlocal effects, involving all the scales,
including the energy pumping scales. In short, this part is infrared
divergent. The vorticity part which we compute, is ultraviolet
divergent,
it it determined by the small scale fluctuations of the vortex sheet,
represented by the Liouville field.

In absence of direct evidence we may try to stretch the rules and
estimate our
intermittency exponents from the moments of velocity. I would expect
this to be
an upper estimate, as intermittency tends to decrease with the
removal of the
large scale effects. In my opinion, there is still no direct evidence
for the
anomalous dimensions in vorticity correlators. It would be most
desirable to
fill this gap in real or numerical experiments.

Let us stress once again, that above speculations do not pretend to
be
a theory of turbulence. Still they may give us an idea how to build
one.

In statistical mechanics the Gibbs distribution was the beginning,
not the
end of the theory. If this approach is correct, which remains to be
seen,
then all the  work is also ahead of us in the string theory of
turbulence.
The string theory methods should be fitted for this unusual case,
and,
perhaps, some scaling and area laws could be established. I appeal to
my
friends in the string community. Look at the turbulence, this is a
beautiful
example of the fractal geometry with extra advantage of being
guaranteed
to exist!

\section{Acknowledgments}

I am grateful to Albert Schwarz for stimulating discussions at
initial
stage of this work and to Herman Verlinde and members of the Santa
Barbara workshop in Quantum Gravity and String Theory for discussion
of the
intermediate results. Final results were discussed with  Vadim Borue,
Andrew
Majda and  Mark Wexler who made various useful comments.This research
was
sponsored in part by the Air Force Office of Scientific Research
(AFSC) under
contract F49620-91-C-0059. The United States Government is authorized
to
reproduce and distribute reprints for governmental purposes
notwithstanding
any copyright notation hereon. This research was supported in part by
the
National Science Foundation under Grant No. PHY89-04035 in ITP of
Santa
Barbara.

\appendix

\section{Poisson brackets for the Euler equation}

One may  derive the Poisson brackets  for the Euler equation by
comparing
the right side of this equation with
\be
\PBR{\val(r), H} = \int d^3 r' \PBR{\val(r),\vbe(r')}
\fbyf{H}{\vbe(r')} =
\int d^3 r' \PBR{\val(r),\vbe(r')} \vbe(r')
\ee

Comparing this with the Euler equation we find
\be
\PBR{\val(r),\vbe(r')}  = T_{\alpha\nu}(r-r') \omega_{\nu\beta}(r') +
\dots
\ee
where dots stand for the $ \dbe $ terms which drop in the above
integral.
These terms should be restored in such a way that the \PB would
become
skew symmetric plus they must be divergenceless in $ r' $ as well as
in $r$
\be
\pp{r_{\alpha}} \PBR{\val(r),\vbe(r')} = \pp{r'_{\beta}}
\PBR{\val(r),\vbe(r')} = 0
\ee
The unique solution is given by the formula in the text.

It is also worth noting that one could derive the \PB for velocity
field
using Clebsch variables, which are known to be an ordinary $ (p,q) $
pair.
The velocity field is represented as an integral
\be
\val(r) = \int d^3 r' T_{\alpha\beta}(r-r') \Phi_1 (r')\dbe
\Phi_2(r')
\ee
and the corresponding vorticity is local
\be
\omega_{\alpha\beta}(r) = e^{ij} \dal \Phi_i \dbe \Phi_j
\ee

The Euler equations are equivalent to the following  equations
\be
\dot{\Phi_i} + \vbe \dbe \Phi_i =0
\label{EQMOT}
\ee
which simply state that the Clebsch fields are passively advected by
the flow.
These equations have an explicit Hamiltonian form
\be
\dot{\Phi}_{i} = - e^{ij}\, \frac{\delta H}{\delta \Phi_{j}}
\label{CLEBSHHAM}
\ee
where it is implied that the Hamiltonian is the same, with velocity
expressed in Clebsch fields.

Now the \PB for the velocity fields can be computed in a standard way
\be
\PBR{\val(r_1),\vbe(r_2)} = \int d^3 r e^{ij} \,
\fbyf{\val(r_1)}{\Phi_i(r)}\fbyf{\vbe(r_2)}{\Phi_j(r)}
\ee
which again yields the formula in the text.
This derivation is not as general as the previous one, since there
are
some flows which cannot be globally described in Clebsch variables.

\section{Helmholtz vs Euler dynamics }

Let us study the variation of the vortex cell velocity field. Our
first
objective would be to show that the time variation according to the
Helmholtz equation
\be
\delta X_{\alpha}(\rho) = dt\, \val(X(\rho))
\ee
reproduces the Euler equation
\be
\delta\val(r) = dt\, \left(\vbe(r) \omega_{\alpha\beta}(r) - \dal
h(r)
\right)
\ee

Simple calculation  of the variation of the initial definition with
integration by parts yields
\be
\delta \val(r) = \left(e_{\alpha\beta\nu} \d_{\gamma}-
e_{\alpha\beta\gamma} \d_{\nu}\right)\d_{\beta}
\int_D  d^3 \rho  \frac{\delta
X_{\nu}(\rho)\Omega_{\gamma}(\rho)}{4\pi|r-X(\rho)|}
\ee
Now we use the identity
\be
e_{\alpha\beta\nu} \d_{\gamma}-e_{\alpha\beta\gamma} \d_{\nu} =
e_{\beta\gamma\nu} \d_{\alpha}-e_{\alpha\gamma\nu} \d_{\beta}
\ee
(the difference between the left and the right sides represents the
completely skew symmetric tensor of the fourth rank in three
dimensions,
which must vanish). This gives us the following two terms in velocity
variation
\be
\delta \val(r) =  -\dal \delta h(r) + e_{\alpha\gamma\nu} \int d^3
\rho\,
\delta X_{\nu}(\rho) \Omega_{\gamma}(\rho) \delta^3(r-X(\rho))
\label{VV}
\ee
where
\be
\delta h(r) = e_{\gamma\beta\nu}\dbe \int d^3 \rho\,
\frac{\delta X_{\nu}(\rho)\Omega_{\gamma}(\rho)}{4\pi|r-X(\rho)|}
\ee

The first term is purely potential, it can be reconstructed from
the second term by solving $ \dal \delta\val(r) = 0 $. Now we see
that the
Helmholtz variation  reproduces the Euler equation with the enthalpy
\be
 h(r) =  e_{\gamma\beta\nu}\dbe \int d^3 \rho\,
\frac{v_{\nu}(X(\rho))\Omega_{\gamma}(\rho)}{4\pi|r-X(\rho)|}
\ee

At the same time we see that the Helmholtz variation is defined
modulo
gauge transformation
\be
\delta X_{\nu}(\rho) \Ra \delta X_{\nu}(\rho) + f(\rho)
\Omega_{\nu}(\rho)
\ee
which leaves the velocity variation invariant.

Note that the functional derivative
\be
\fbyf{ \val(r)}{X_{\nu}(\rho)} = \left(e_{\alpha\beta\nu}
\d_{\gamma}-
e_{\alpha\beta\gamma} \d_{\nu}\right)\d_{\beta}
 \frac{\Omega_{\gamma}(\rho)}{4\pi|r-X(\rho)|}
\ee
is a traceless tensor in $ \alpha, \nu $ . This is sufficient for the
volume conservation
\be
\dt (DX) \propto \int d^3 \rho \fbyf{\val(X(\rho))}{X_{\alpha}(\rho)}
= 0
\ee
which is  the Liouville theorem for the vortex dynamics.

We found the following Poisson bracket for the $ X $ field
\be
\PBR{X_{\alpha}(\rho), X_{\beta}(\rho')} = -\delta^3(\rho-\rho')
e_{\alpha\beta\gamma}
\frac{\Omega_{\gamma}(\rho)}{\Omega_{\mu}^2(\rho)}
\ee
The equations of motion corresponding to these \PB read
\be
\dt X_{\alpha}(\rho) = \PBR{X_{\alpha}(\rho), H} =
\int d^3 \rho' \PBR{X_{\alpha}(\rho), X_{\beta}(\rho')}
\fbyf{H}{X_{\beta}(\rho')}
\ee

Let us compare these equations with the usual Helmholtz dynamics.
The variation of the Hamiltonian reads
\be
\delta H = \int d^3 r  \val(r) \delta \val(r)
\ee
Substituting here the velocity variation \rf{VV} we could drop the $
\dal
\delta h $ term as it vanishes after integration by parts.
As a result we find
\be
\fbyf{H}{X_{\alpha}(\rho)} = e_{\alpha\beta\gamma}
\vbe\left( X(\rho)\right) \Omega_{\gamma}(\rho)
\label{GenH}
\ee

Finally,  in the equation of motion we have
\bea
\dt X_{\alpha}(\rho) =
-\int d^3 \rho' \delta^3(\rho-\rho')
e_{\alpha\beta\gamma}
\frac{\Omega_{\gamma}(\rho)}{\Omega_{\mu}^2(\rho)}
e_{\beta\mu\nu}  v_{\mu}\left( X(\rho)\right) \Omega_{\nu}(\rho) \br
= \left(\delta_{\alpha\beta} -
\frac{\Omega_{\alpha}(\rho)\Omega_{\beta}(\rho)}{\Omega_{\mu}^2(\rho)
}\right)
\vbe\left( X(\rho)\right)
\eea

So, the Poisson brackets correspond to the motion in {\em
transverse direction } to the gauge transformations. Or, to put it in
different terms, we could modify the Helmholtz dynamics by adding the
time
dependent gauge transformations $ \delta \rho(t) $ to the time shift
of
the vortex
sheet
\be
\dt X_{\alpha}(\rho) = \val\left(X(\rho)\right) + \d_a
X_{\alpha}(\rho)
\dt \delta \rho^a
\ee
where
\be
\dt \delta \rho^a = -
\frac{\Omega^a(\rho)\Omega_{\beta}(\rho)\vbe\left( X(\rho)\right)
}{\Omega_{\mu}^2(\rho)}
\ee

Should we insist on the unmodified Helmholtz dynamics, we would have
to
admit that this cannot be achieved by any Poisson brackets. This is
so
called generalized Hamiltonian dynamics, where the formula \rf{GenH}
for the variation of the Hamiltonian cannot be uniquely solved for
the
velocity. The terms corresponding to the gauge transformation remain
unspecified. In our opinion, this difference is immaterial, as the
Helmholtz dynamics is indistinguishable from the one with the Poisson
brackets.

\end{document}